# Dados abertos governamentais no contexto de Políticas Públicas de Saúde e Sistemas Prisionais: Realidade ou Utopia?

OPEN GOVERNMENT DATA [1] IN THE CONTEXT OF PUBLIC HEALTH AND PRISON SYSTEMS: REALITY OR UTOPIA?

*Rafael Antônio Lima Cardoso[1]*
*Glauco de Figueiredo Carneiro[2]*
*José Euclimar Xavier de Menezes[3]*

**1.** MESTRANDO EM SISTEMAS E COMPUTAÇÃO /UNIVERSIDADE SALVADOR (UNIFACS).
http://lattes.cnpq.br/2712724638701758

**2.** DOUTOR EM CIÊNCIA DA COMPUTAÇÃO / UNIVERSIDADE SALVADOR (UNIFACS).
http://lattes.cnpq.br/4951846457502161

**3.** PHD EM FILOSOFIA, UNISBA/UNIFACS. PÓS DOUTOR/U.LATERANENSI/ROMA; PÓS DOUTOR UNIVERSIDADE DE SALAMANCA/ESPANHA.
http://lattes.cnpq.br/5194408237403794
https://orcid.org/0000-0001-7839-7931

## RESUMO

Atualmente, pode-se constatar a prática da transparência no acesso e uso de dados abertos governamentais para diversas finalidades. Esta prática revela um importante requisito para a implementação da governança participativa. A literatura relata um conjunto mínimo de critérios que devem ser atendidos para que determinado repositório de dados governamentais seja considerado aberto. Este artigo apresenta avaliações de alinhamento de dados governamentais nacionais e internacionais a estes critérios. Os repositórios selecionados para análise focam em políticas públicas de saúde e sistemas prisionais. Os resultados apresentam diferentes níveis de alinhamento dos portais analisados aos critérios de referência. Pode-se concluir que alguns dos portais colaboram para que as práticas de dados abertos governamentais sejam consideradas uma realidade, enquanto outros indicam que ainda existe um longo caminho para a implementação de oportunidades de melhorias.

**Palavras-chave**: Dados Abertos Governamentais. Políticas Públicas de Saúde. Sistemas Prisionais.

## ABSTRACT

There are many initiatives of transparency reported in the access and use of government open data

---

65

for different purposes. This practice reveals an important requirement to accomplish the participatory governance. The literature has reported a minimal set of criteria to categorize a specific data repository as open. This paper discusses to which extent specific national and international data repositories address these criteria. The analyzed repositories focus on public health and prison systems. The results show different levels of alignment to the criteria and provide evidence that the adoption of government open data practices are a reality. On the other hand, there is still a long way to achieve full alignment to the stated criteria.

**Key-words:** Open Government Data. Public Health Policies. Prison Systems.

# INTRODUÇÃO

Iniciativas de dados abertos governamentais têm potencial para empoderamento da sociedade de forma que haja acesso pleno e livre aos dados (ALAOUIE, 2019). O acesso a dados públicos promove o engajamento do cidadão a temas de interesse da sociedade, dentre eles, dados relacionados a políticas públicas de saúde e sistemas prisionais. Existem três razões principais para a promoção de dados abertos governamentais (ATTARRD, 2015; DATA OPEN GOVERNMENT): a *transparência*, o valor *social/comercial dos dados*, e a *governança participativa*. A transparência é um dos pilares fundamentais da democracia, sendo exercitada ao se promover o acesso do cidadão à informação governamental (DA SILVA, ET AL. 2014). A transparência pode proporcionar as condições para que o cidadão e entidades, sejam estas governamentais ou não, possam monitorar e acompanhar as iniciativas, demandas e resultados do governo. A transparência não se limita a simplesmente conceder acesso aos dados, mas também permitir o seu uso intensivo, como também reuso e distribuição. Em relação ao *valor social e comercial associado aos dados*, os governos produzem e coletam grande volume de dados de diferentes domínios (ALEXOPOULOS, ET AL. 2014), podendo ser utilizados para diferentes finalidades que podem viabilizar o seu uso de forma inovadora tanto para o cidadão como para as diversas organizações que compõem a sociedade. A publicação de dados governamentais estabelece uma das vias de comunicação entre a sociedade e o governo (DA SILVA, ET AL. 2014) e, portanto, traz oportunidades para que o cidadão participe ativamente do processo de governança, podendo este acompanhar as iniciativas das entidades governamentais em suas diversas frentes de atuação (Alexopoulos, et al. 2014). A disponibilização destes dados ocorre principalmente por intermédio de sites e portais online. Através de repositórios online, tanto a sociedade, entidades governamentais e pesquisadores realizam consultas, *e análise* dos dados de interesse.

Este artigo analisa o grau de alinhamento de um conjunto específico de bases de dados públicas nacionais e internacionais com foco em dados sobre políticas públicas de saúde e sistemas prisionais. A referência para a análise do grau de alinhamento é a comparação contra uma lista de princípios e critérios reconhecidos internacionalmente: os Oito Princípios de Dados Abertos Governamentais (OPEN DATA GOVERNMENT, 2007), os Sete Princípios adicionais de Dados Abertos Governamentais (OPEN DATA GOVERNMENT, 2007) e a classificação Cinco Estrelas proposta por Tim Bernes-Lee (2007).

O restante deste artigo é organizado da seguinte forma. A seção 2 aborda o conceito de dados abertos governamentais e iniciativas nacionais neste tema. A seção 3 apresenta os repositórios nacionais e internacionais de dados abertos selecionados. Na seção 4 avaliamos o grau de alinhamento com os princípios e critérios de qualidade para dados abertos dos repositórios selecionados. A seção 5 apresenta a conclusão do que foi apresentado no artigo.

## DADOS ABERTOS GOVERNAMENTAIS

De acordo com o Artigo 2o do decreto no 8.777 de 11 de maio de 2016 do Governo Federal do Brasil (PRESIDÊNCIA DA REPÚBLICA), *dados* são uma sequência de símbolos ou valores, representados em qualquer meio, produzidos como resultado de um processo natural ou artificial. Ainda de acordo com o mesmo decreto, *dados abertos* são acessíveis ao público, representados em meio digital, estruturados em formato aberto, processáveis por máquina, referenciados na internet e disponibilizados sob licença aberta que permita sua livre utilização, consumo ou cruzamento, limitando-se a creditar a autoria ou a fonte. Pelo fato de serem referenciados na internet, podem ser localizados facilmente e, sendo estruturados em formato aberto e processáveis por máquina, podem também ser utilizados recursos computacionais para seu processamento para diferentes finalidades. Por estes motivos, não podem estar sujeitos a patentes ou quaisquer mecanismos de controle.

Segundo o Open Knowledge Foundation (OPEN KNOWLEDGE FOUNDATION S.D.), *dados abertos* podem ser utilizados, reutilizados e disponibilizados livremente por qualquer pessoa física ou jurídica, desde que sejam respeitadas as questões de privacidade. A

**Olhares das ciências sobre as questões sociais**

partir das características apresentadas, tem-se que os dados governamentais têm por vocação natural serem também *abertos*, daí o termo *dados abertos governamentais*. Dentre as iniciativas da administração pública brasileira para a implementação de dados abertos: Lei de Acesso à Informação Pública (LEI NO. 12.527, de 18 de novembro de 2011), a Parceria para Governo Aberto (OPEN GOVERNMENT PARTNERSHIP - OGP), a criação da infraestrutura Nacional de Dados Abertos (INFRA ESTRUTURA DE DADOS ABERTOS/INDA).

Existem pelo menos três diferentes tipos de perfis de ferramentas que podem ser utilizadas para acesso, uso e exploração de dados abertos governamentais: ferramentas para exploração de dados, ferramentas analíticas e ferramentas para a criação de *mashups* (ASSAF, TRONCY E SENART 2015). As ferramentas de exploração de dados podem servir para reduzir as possíveis barreiras técnicas que os usuários possam vir a enfrentar no acesso e uso aos dados. Para esta finalidade, são oferecidas funcionalidades de visualização e exploração de dados (COLPAERT, ET AL. 2013). Ferramentas analíticas incluem recursos para a visualização de dados como um mecanismo para apoiar a compreensão e a comunicação de grandes volumes de dados. Isto torna o seu uso mais efetivo para o cidadão, apoiando na identificação de padrões que de outra forma não seriam facilmente descobertos em repositórios publicados (GRAVES E HENDLER, 2013). Finalmente, as ferramentas de *mashups* possibilitam ao cidadão juntar diferentes conjuntos de dados, o que pode promover novas possibilidades para a sua exploração (CHATFIELD E REDDICK, 2017).

Outras ferramentas também foram discutidas na literatura como alternativas para acesso e uso de dados abertos governamentais. Dentre estas, tem-se a Application Programming Interface/API (ALEXOPOULOS, ET AL, 2014; GRAVES E HENDLER 2013). Uma API é um conjunto de rotinas (programas) disponíveis para o desenvolvimento de software e um meio efetivo para que sejam acessados e obtidos dados abertos governamentais, incluindo a possibilidade de construir representações visuais para estes dados (GRAVES E HENDLER, 2013). Além das APIs, trabalhos anteriores também identificaram ferramentas de conversão de formato de dados (ATTARD, ET AL, 2015). Estas ferramentas viabilizam a criação de arquivos de diferentes formatos para os mesmos dados disponibilizados no repositório, o que aumenta a disponibilidade destes dados para atividades analíticas (POPOVIC, KEZUNOVIC E KRSTAJIC, 2015). Além disso, vale ressaltar que portais de dados abertos, como aqueles que utilizam a plataforma de gerenciamento de dados

**Olhares das ciências sobre as questões sociais**

CKAN, disponibilizam painéis para representação visual de sumários estatísticos a respeito dos dados armazenados. Dois exemplos de itens que podem ser representados nestes sumários são: (1) os repositórios mais acessados pelos usuários do portal e (2) quais os temas relacionados aos principais repositórios disponíveis.

Dados abertos governamentais podem ser qualificados de acordo com requisitos e características específicos, amplamente conhecidos como os *Oito Princípios dos Dados Abertos Governamentais* (OPEN DATA GOVERNMENT, 2007), conforme apresentado na Tabela 1.

Tabela 1 - Oito Princípios dos Dados Abertos Governamentais

| **Princípios** | **Descrição** |
|---|---|
| (1) Completo | Todos os dados públicos devem ser disponibilizados. Dados públicos são dados que não estão sujeitos a restrições de privacidade, segurança ou privilégios de acesso. |
| (2) Primário | Os dados devem ser coletados na fonte com o maior nível de detalhamento possível, e não de forma agregada ou modificada. |
| (3) Atualizado | Sua disponibilidade deve ser feita tão rapidamente quanto necessário para preservar o valor dos dados. |
| (4) Acessível | Os dados devem estar disponíveis para a mais ampla gama de usuários e as mais diversas finalidades. |
| (5) Processável por Máquinas | Os dados devem ser razoavelmente estruturados de modo a permitir o processamento automatizado. |
| (6) Não Discriminatório | Os dados devem estar disponíveis para qualquer pessoa, sem necessidade de registro. |
| (7) Não Proprietário | Os dados devem estar disponíveis em um formato sobre o qual nenhuma entidade tem o controle exclusivo. |
| (8) Licença Livre | Os dados não estão sujeitos a quaisquer direitos de autor, patentes, marcas comerciais ou regulamento secreto. Pode ser permitida uma razoável privacidade e restrições de privilégio e segurança. |

Além dos princípios já mencionados, existem sete princípios adicionais para dados abertos governamentais (OPEN DATA GOVERNMENT, 2007), estendendo os critérios e boas práticas que devem ser adotadas pelos governos para disponibilização de dados. Na tabela 2 são apresentados estes princípios adicionais.

**Olhares das ciências sobre as questões sociais**

Tabela 2 - Sete Princípios adicionais dos Dados Abertos Governamentais

| Princípios | Descrição |
|---|---|
| (9) Online e sem custo | Os dados são disponibilizados sem custo, tanto para acesso pelo site, como para extração e download. |
| (10) Permanente | Os dados devem ser mantidos ao longo do tempo (histórico). |
| (11) Confiável | Os dados publicados devem ser assinados digitalmente, ou atestar a data de publicação, criação, autenticidade e integridade. |
| (12) A Presunção de Abertura | Baseia-se na liberdade de acesso e uso da informação e na forma como os registros e catálogos de dados são gerenciados. |
| (13) Documentado | A documentação sobre o formato e o significado dos dados aumenta a efetividade do seu uso. |
| (14) Seguro para Acesso e Uso | Os dados devem ser publicados usando formatos dados que não incluam conteúdo executável(evitar vírus e worms). |
| (15) Participação da Sociedade | A Contribuição do público é essencial para disseminar informações de maneira que elas tenham valor. |

Além dos quinze princípios já apresentados, existem outros critérios de qualidade aplicados aos dados abertos, sendo o mais destacado destes o proposto por Tim Berners-Lee (TIM BERNERS-LEE, 2007) para o estabelecimento de cinco níveis de implementação de dados abertos conhecidos por *Cinco Estrelas* e apresentados na Tabela 3. Os níveis são acumulativos, ou seja, para que um nível seja atendido, tem-se como pré-requisito que os níveis inferiores também devam ser atendidos. Verifica-se que existem relacionamentos de equivalência entre os três primeiros níveis de estrelas com os princípios *(4) Acessível*, *(5) Processável por Máquina* e *(7) Não proprietário*, respectivamente, já apresentados na Tabela 1. Por este motivo, este três primeiros níveis das Estrelas referenciados na Tabela 3 não foram considerados nas Tabelas 5 e 6 da Seção 4 que apresentam a análise do grau de alinhamento aos princípios reconhecidos de dados abertos governamentais.

Tabela 3 - Níveis de qualidade para dados abertos

| Níveis | Descrição |
|---|---|
| Uma Estrela | Os dados são disponibilizados pela internet independente do formato. (Similar ao princípio *(4) Acessível* da Tabela 1). |
| Duas Estrelas | Os dados são estruturados como arquivos tabulares (Similar ao princípio *(5) Processável por Máquina* da Tabela 1). |
| Três Estrelas | Os dados são disponibilizados em formatos não proprietários, por exemplo, *(ex.: csv)* (Similar ao princípio *(7) Não proprietário* da Tabela 1 ). |
| Quatro Estrelas | Os dados devem possuir URIs para facilita seu acesso. |
| Cinco Estrelas | Os dados devem possuir ligações com outros dados. Governos podem possuir links para dados de outros governos ou outras fontes de dados abertos. |

# REPOSITÓRIOS DE DADOS SELECIONADOS

Os resultados da avaliação de aderência a processos e produtos gerados a partir de dados abertos não são úteis somente às organizações e pessoas diretamente envolvidas no tema, mas também a pesquisadores, jornalistas, cidadãos e demais interessados. Os tomadores de decisão e aqueles diretamente envolvidos na disponibilização de dados abertos podem fazer uso dos resultados para monitorar o progresso em relação a objetivos pré-estabelecidos e também para identificar áreas e práticas que precisam ser melhoradas. Aos demais interessados, os resultados disponíveis publicamente também permitem comparar as diferentes iniciativas, portais e bases de dados, dando oportunidade para a identificação de boas práticas que se destacarem. Por último, os resultados também são relevantes para a responsabilização, tendo em vista que permitem que governos e agências sejam responsabilizados pelas suas decisões, atividades e alocação dos recursos (VANCAUWENBERGHE, 2018).

Uma pesquisa foi conduzida para identificar um conjunto representativo de repositórios nacionais e internacionais com dados relacionados a políticas públicas de saúde e sistemas prisionais. Para os portais relacionados a sistemas prisionais, adotou-se como estratégia a consulta a profissionais e especialistas que atuam em atividades relacionadas a este tema para que pudessem descrever quais seriam os repositórios de referência de pesquisa disponíveis para consulta por parte daqueles que atuam tanto na gestão, na fiscalização, como também na atuação jurídica no sistema prisional. Os portais e repositórios identificados são listados a seguir. O primeiro portal indicado foi o Sistema Eletrônico de Execução Unificado (SEEU), mantido pelo Conselho Nacional e que centraliza e uniformiza a gestão de processos de execução penal em todo o país. O SEEU possibilita acesso simultâneo concedido a diferentes perfis de usuários, tais como promotores de justiça, defensores públicos, advogados, gestores prisionais entre outros (CONSELHO NACIONAL DE JUSTIÇA). Face à importância dos procedimentos de execução penal sobre o tema sistemas prisionais, tem-se a expectativa que os dados relacionados às execuções, em suas diversas etapas e possibilidades, estejam disponíveis para observação analítica pelos diversos interessados. Por exemplo, a expectativa de ingresso e também de soltura de internos no sistema prisional pode ser estimada por intermédio da análise do número global de processos em execução cujas condições indiquem o encaminhamento ou liberação da carceragem em uma dada região.

**Olhares das ciências sobre as questões sociais**

Entretanto, tal análise será possível somente caso os dados possam ser analisados em sua totalidade pelo interessado ou caso já existe filtro específico para esta funcionalidade no referido portal. Este exemplo mostra a flexibilidade promovida ao interessado quando os dados são disponibilizados na sua totalidade para análise de acordo com suas necessidades específicas, tendo em vista que não será possível ao provedor dos dados prever todas as possibilidades de uso para implementação específico de filtros para esta finalidade. Esta é uma oportunidade para a verificação do grau de alinhamento destes dados em relação aos princípios selecionados e já apresentados neste artigo. Já o Banco Nacional de Monitoramento de Prisões (BNMP 2.0) é um sistema eletrônico que auxilia as autoridades judiciárias da justiça criminal na gestão de documentos atinentes às ordens de prisão/internação e soltura expedidas em todo o território nacional, materializando um Cadastro Nacional de Presos (CONSELHO NACIONAL DE JUSTIÇA-SISTEMA CARCERÁRIO). O Conselho Nacional do Ministério Público (CNMP) dispõe de repositório de dados alinhado à Resolução Nº 56, de 22 de junho de 2010. Se a análise dos dados do SEEU pode fornecer uma expectativa do encarceramento e soltura no sistema prisional, dentre outras informações, o INFOPEN fornece os dados concretos destes eventos, havendo a possibilidade de se correlacionar os dados fornecidos pelas duas bases de dados no que tange à expectativa versus realização de encarceramento e soltura em uma ou mais unidades do sistema prisional. De forma similar ao cenário descrito anteriormente, este será possível somente se os dados estiverem disponíveis para a implementação desta correlação. O INFOPEN é um sistema de informações estatísticas do sistema penitenciário brasileiro, atualizado pelos gestores dos estabelecimentos desde 2004, sintetiza informações sobre os estabelecimentos penais e a população prisional (MINISTÉRIO DA JUSTIÇA E SEGURANÇA PÚBLICA). O GEO Presídios apresenta dados dos relatórios do Cadastro Nacional de Inspeções nos Estabelecimentos com o objetivo de consolidar em um único banco de dados as informações sobre as inspeções em todo território nacional, o que viabiliza o controle das inspeções pelos órgãos judiciais (CONSELHO NACIONAL DE JUSTIÇA). Utilizamos três repositórios internacionais na análise. O primeiro deles, o Prison Policy Iniciative, é uma iniciativa de políticas públicas prisionais, sem fins lucrativos que contam com a participação de profissionais da área jurídica que disponibiliza dados, estudos e estatísticas sobre criminalidade e sua relação com sistemas prisionais. O segundo é o Open Canada (GOVERNMENT OF CANADA), uma iniciativa do Governo Canadense aos princípios de dados abertos governamentais que, dentre outros temas, disponibiliza dados relacionados a sistemas prisionais daquele país. O terceiro repositório é por fim o repositório de dados abertos do estado de Nova York[2], que apresenta um amplo conjunto de dados governamentais e de saúde pública, além de possuir um alto grau de alinhamento com os princípios e níveis de qualidade de dados abertos governamentais.

## ANÁLISE DO ALINHAMENTO AOS CRITÉRIOS DE DADOS ABERTOS GOVERNAMENTAIS

O grau de alinhamento dos repositórios selecionados aos critérios de dados abertos governamentais é o foco principal desta seção. A análise foi conduzida em repositórios cujos dados estão diretamente relacionados a sistemas prisionais e saúde pública. A Tabela 5 apresenta os repositórios com dados dos sistemas prisionais, enquanto que a Tabela 6 foca nos dados de saúde pública. Para facilitar a representação do grau de alinhamento aos princípios ou critérios de referência, a cor verde indica alinhamento pleno ao princípio ou critério, a cor vermelha indica o não alinhamento, enquanto que a cor amarela indica alinhamento parcial ao princípio ou critério de referência. A Tabela 4 apresenta as perguntas utilizadas para avaliar objetivamente o grau de alinhamento de cada repositório com os princípios e critérios já apresentados nas tabelas 1, 2 e 3.

---

2 https://www.ny.gov/

Tabela 4 - Descritivo das perguntas aplicadas na análise dos repositórios

| Conceitos | Perguntas da Avaliação | Descrição |
|---|---|---|
| Completo | Pergunta 1.1 | O conjunto de dados disponibilizado é suficiente para alcançar os objetivos estabelecidos de análise? |
| | Pergunta 1.2 | Os dados são originais e atendem questões de segurança e privacidade? |
| Primário | Pergunta 2.1 | Os dados são originais e foram coletadas de fontes primárias? |
| | Pergunta 2.2 | Os dados possuem um nível de detalhamento que atendem os objetivos estabelecidos de análise? |
| | Pergunta 2.3 | A forma como os dados foram coletados é descrita? |
| Atualizado | Pergunta 3.1 | Os dados são atualizados com periodicidade conhecida? |
| Acessível | Pergunta 4.1 | Os dados estão disponíveis na internet? |
| | Pergunta 4.2 | O acesso aos dados é intuitivo? |
| | Pergunta 4.3 | A extração e coleta dos dados pode ser realizada através de API ou Web Services disponibilizada pelo portal? |
| Processável por Máquinas | Pergunta 5.1 | Os arquivos são disponibilizados em formatos conhecidos, e são processáveis por máquina (ex.: CSV, XML, XLS)? |
| | Pergunta 5.2 | Os Metadados são disponibilizados, e as instruções de como utilizá-los? |
| Não Discriminatório | Pergunta 6.1 | O acesso aos dados é livre para qualquer pessoa, sem necessidade de identificação ou registro? |
| Não Proprietário | Pergunta 7.1 | Os dados estão disponibilizados em formatos não proprietários como (CSV, XML, PDF)? |
| Licença Livre | Pergunta 8.1 | Os dados não estão sujeitos a nenhuma regulamentação de direitos autorais, patente, ou marca comercial? |
| Permanente | Pergunta 9.1 | Os dados são disponibilizados juntamente com registros de versionamento? |
| Online e sem custo | Pergunta 10.1 | Os dados são disponibilizados sem custo, tanto para acesso pelo site, como para extração e download? |
| Confiável | Pergunta 11.1 | Os dados publicados são assinados digitalmente, ou é possível atestar a data de publicação, criação, sua autenticidade e integridade? |
| A presunção da Abertura | Pergunta 12.1 | Os dados são disponibilizados seguindo procedimentos definidos em Leis de acesso a informação ou ferramentas adequadas para abertura de dados? |
| Documentado | Pergunta 13.1 | Existe documentação sobre o formato e o significado dos dados(metadados)? |

| Seguro para Acesso | Pergunta 14.1 | Os dados publicados usam formatos que não incluam conteúdo executável evitando vírus e worms? |
|---|---|---|
| Participação da Sociedade | Pergunta 15.1 | São disponibilizados mecanismos para que a sociedade possa acessar, sugerir, reclamar sobre os dados disponibilizados? |

Tabela 5 - Análise de Dados Abertos Governamentais - Dados de Sistemas Prisionais

| Princípios | Perguntas da Avaliação | Repositórios de Dados de Sistema Prisionais | | | | | | |
|---|---|---|---|---|---|---|---|---|
| | | BNMP | SEEU | Prison Policy | Open Canada | CNMP | INFOPEN | Geo Presídios |
| 1- Completo | Pergunta 1.1 | | | | | | | |
| | Pergunta 1.2 | | | | | | | |
| 2- Primário | Pergunta 2.1 | | | | | | | |
| | Pergunta 2.2 | | | | | | | |
| | Pergunta 2.3 | | | | | | | |
| 3- Atualizados | Pergunta 3.1 | Constante | Constante | Anual | Mensal | Anual | Semestral | Mensal |
| 4- Acessível | Pergunta 4.1 | | | | | | | |
| | Pergunta 4.2 | | | | | | | |
| | Pergunta 4.3 | Com Autorização | | | | | | |
| 5- Processável por Máquina | Pergunta 5.1 | CSV, e XLSX | | XLSX | CSV, e XLSX | | | |
| | Pergunta 5.2 | | | | | | | |
| 6- Não Discriminatório | Pergunta 6.1 | | | | | | | |
| 7- Não Proprietário | Pergunta 7.1 | | Web (HTML) | PDF | CSV | PDF e PNG | PDF | PNG, JPG, PDF e SVG |
| 8- Licença Livre | Pergunta 8.1 | | | | | | | |
| 9- Permanente | Pergunta 9.1 | | | | | | | |
| 10- Sem Custo | Pergunta 10.1 | | | | | | | |
| 11- Confiável | Pergunta 11.1 | | | | | | | |
| 12- A presunção da Abertura | Pergunta 12.1 | | | | | | | |
| 13- Documentado | Pergunta 13.1 | | | | | | | |

| 14- Seguro para Acesso | Pergunta 14.1 | 🟢 | 🟢 | Apenas XLSX | 🟢 | 🟢 | 🟢 | 🟢 |
| 15- Participação da Sociedade | Pergunta 15.1 | 🟢 | 🟢 | 🟢 | 🟢 | 🟢 | 🟢 | 🔴 |
| Qualidade de Dados Abertos | Níveis | 3 Estrelas | 1 Estrela | 2 Estrelas | 3 Estrelas | 1 Estrela | 1 Estrela | 1 Estrela |

Tabela 6 - Análise de Dados Abertos Governamentais - Dados de Saúde Pública

| Princípios | Perguntas da Avaliação | Repositórios de Dados de Saúde | |
| --- | --- | --- | --- |
| | | ny.gov | DATASUS |
| 1- Completo | Pergunta 1.1 | 🟢 | 🔴 |
| | Pergunta 1.2 | 🟢 | 🔴 |
| 2- Primário | Pergunta 2.1 | 🟢 | 🔴 |
| | Pergunta 2.2 | 🟢 | 🔴 |
| | Pergunta 2.3 | 🟢 | 🔴 |
| 3- Atualizados | Pergunta 3.1 | Constante | 🔴 |
| 4- Acessível | Pergunta 4.1 | 🟢 | 🔴 |
| | Pergunta 4.2 | 🟢 | 🔴 |
| | Pergunta 4.3 | 🟢 | 🔴 |
| 5- Processável por Máquina | Pergunta 5.1 | CSV, XLSX, XML e JSON | DBC |
| | Pergunta 5.2 | JSON | 🔴 |
| 6- Não Discriminatório | Pergunta 6.1 | 🟢 | 🟢 |
| 7- Não Proprietário | Pergunta 7.1 | 🟢 | 🔴 |
| 8- Licença Livre | Pergunta 8.1 | 🟢 | 🟢 |
| 9- Permanente | Pergunta 9.1 | 🟢 | 🟢 |
| 10- Sem Custo | Pergunta 10.1 | 🟢 | 🟢 |
| 11- Confiável | Pergunta 11.1 | 🟢 | 🟢 |
| 12- A presunção da Abertura | Pergunta 12.1 | 🟢 | 🟡 |
| 13- Documentado | Pergunta 13.1 | 🟢 | 🔴 |
| 14- Seguro para Acesso | Pergunta 14.1 | 🟢 | 🟢 |
| 15- Participação da Sociedade | Pergunta 15.1 | 🟢 | 🔴 |
| Qualidade de Dados Abertos | Níveis | 5 Estrelas | 1 Estrela |

As observações apresentadas a seguir refletem os resultados apresentados nas Tabelas 5 e 6. Além disso, pode-se verificar nestas tabelas a situação de cada princípio ou critério avaliado para cada repositório: 21 perguntas da Tabela 4 e

**Olhares das ciências sobre as questões sociais**

duas perguntas dos níveis de qualidade 4 e 5 estrelas da Tabela 3, totalizando 23 itens de avaliação. Apenas os critérios de qualidade 4 e 5 foram utilizados, visto que os três primeiros critérios possuem uma equivalência com três princípios já mencionados na Tabela 1 da Seção 2, não sendo necessário, portanto, incluí-los na avaliação. Pode-se verificar na Tabela 7 que o repositório do Estado de Nova York apresenta aderência de 100\% aos critérios, enquanto que, no outro extremo, o repositório do DataSUS apresenta o menor grau de aderência, somente 39,1\%.

Tabela 7 – Grau de Alinhamento aos Princípios e Critérios

| Repositório | País | Grau de Alinhamento (%) |
|---|---|---|
| NY.gov | Estados Unidos | 100 (23 de 23) |
| Open Canada | Canadá | 91.3 (21 de 23) |
| BNMP | Brasil | 71.7 (16.5 de 23) |
| INFOPEN | Brasil | 67.3 (15.5 de 23) |
| CNMP | Brasil | 67.3 (15.5 de 23) |
| Prison Policy | Estados Unidos | 58,6 (13.5 de 23) |
| Geo Presídios | Brasil | 54,3 (12.5 de 23) |
| SEEU | Brasil | 58,6 (13.5 de 23) |
| DATASUS | Brasil | 39,1 (9 de 23) |

A maioria dos repositórios nacionais não disponibiliza APIs, fato que torna estes repositórios com recursos limitados para pesquisa e coleta de forma automatizada. Além disso, a ausência de API não permite a interoperabilidade com outros repositórios, dificultando o uso de técnicas analíticas em sua plenitude como requisito relevante para a obtenção de indicadores de apoio à tomada de decisão. Os repositórios nacionais disponibilizam também na sua maioria seus dados em formatos não processáveis por máquinas. Boa parte deles são disponibilizados em formatos de imagem *(jpg, png)*, exigindo que a extração ou coleta demande mais esforço quando comparado ao requerido para tratamento de dados representados em formatos processáveis por máquina. O argumento para estes casos é que a disponibilização de formatos de dados processáveis facilita e torna mais efetivo o trabalho dos potenciais interessados nestes dados. Outro item importante no grau de alinhamento é a disponibilização de metadados que atuam como descritores dos dados disponibilizados no que tange ao conteúdo dos campos e seus respetivos tipos de dados. Por exemplo, os metadados fornecem a descrição semântica dos dados apresentados em determinado campo, assim como o seu formato, seja ele numérico, texto, booleano, dentre outros. Sem o auxílio dos metadados, os dados tornam-se menos intuitivos. Os metadados são utilizados para criar os dicionários de dados, tornando claro o seu significado. Os catálogos de dados utilizam como base os metadados para apresentar o conteúdo de um repositório consultado, apoiando a localização dos dados através de mecanismos de pesquisa na internet. A maioria do repositórios não apresenta explicitamente licenças de domínio público, ou estão vinculados aos próprios órgãos governamentais, não ficando claro se existem limitações sobre o uso dos dados disponibilizados.

Observamos que o repositório nacional do DataSUS que tem papel de destaque para o acompanhamento dos dados do Sistema Único de Saúde pelos diversos atores deste processo, incluindo os prestadores de serviço, possui um baixo grau de alinhamento, podendo impactar na baixa qualidade dos dados disponibilizados. Isto certamente poderá afetar a experiência na pesquisa, acesso e coleta dos dados por parte do interessados. Os repositórios nacionais com dados dos sistemas prisionais tem um grau de alinhamento maior que os de saúde pública. Verifica-se que os repositórios nacionais avaliados ainda estão abaixo do ideal, demonstrando uma realidade ainda distante dos padrões internacionais para disponibilizar dados abertos governamentais. Em contrapartida, os repositórios internacionais apresentam um alto grau de alinhamento, o que pode refletir na qualidade dos dados diponibilizados e na rica experiência no acesso, pesquisa e coleta.

Apesar de os portais e os sites nacionais apresentarem baixo alinhamento aos princípios e qualidade dos dados abertos, verifica-se que os portais internacionais possuem maior nível de aderência a estes princípios, indicando, portanto, a viabilidade da disponibilização de dados com qualidade e dentro de padrões estabelecidos mundialmente. Constata-se o desafio em demonstrar a importância deste alinhamento para as entidades governamentais, e adequar as estruturas já existentes para esta finalidade.

## CONCLUSÃO

Diante das evidências comparadas apresentadas, observamos que existe uma demanda crescente por dados abertos, e a qualidade destes dados, assim como sua disponibilidade, impactam no consumo dos mesmos. As entidades governamentais, de modo geral, têm disponibilizado dados, mesmo que parcialmente, sob sua gestão. Entretanto, verifica-se que não são seguidos os princípios e critérios reconhecidos internacionalmente para dados abertos governamentais, fato que pode aumentar o esforço requerido e até mesmo dificultar a busca, utilização e interoperabilidade destes dados. O alinhamento dos governos em relação aos princípios e ferramentas tecnológicas específicas para disponibilização dos dados, contribuiria para uma estrutura mais homogênea, facilitando seu acesso e utilização pelos interessados.

# REFERÊNCIAS

**Olhares das ciências sobre as questões sociais**